\documentclass[twocolumn]{aastex631}

\begin{document}

\title{\textit{DECaPS} and \textit{SMA} discovery of a highly-inclined class~I YSO with an outflow: IRAS~08235-4316}

\author[0000-0002-4248-5443]{Joshua Bennett Lovell}
\affiliation{Center for Astrophysics, Harvard \& Smithsonian, 60 Garden Street, Cambridge, MA 02138-1516, USA}

\author[0000-0002-5688-6790]{Kristina Monsch}
\affiliation{Center for Astrophysics, Harvard \& Smithsonian, 60 Garden Street, Cambridge, MA 02138-1516, USA}

\author[0000-0002-3490-146X]{Garrett K. Keating}
\affiliation{Center for Astrophysics, Harvard \& Smithsonian, 60 Garden Street, Cambridge, MA 02138-1516, USA}

\author[0000-0003-1526-7587]{David J. Wilner}
\affiliation{Center for Astrophysics, Harvard \& Smithsonian, 60 Garden Street, Cambridge, MA 02138-1516, USA}

\author[0000-0003-3122-4894]{Gordian Edenhofer}
\affiliation{Center for Astrophysics, Harvard \& Smithsonian, 60 Garden Street, Cambridge, MA 02138-1516, USA}
\affiliation{Max Planck Institute for Astrophysics, Karl-Schwarzschild-Str. 1, 85741 Garching, Germany}
\affiliation{Ludwig Maximilian University of Munich, Geschwister-Scholl-Platz 1, 80539 Munich, Germany}

\author[0000-0003-0685-3621]{Mark Gurwell}
\affiliation{Center for Astrophysics, Harvard \& Smithsonian, 60 Garden Street, Cambridge, MA 02138-1516, USA}

\author[0000-0002-1407-7944]{Ramprasad Rao}
\affiliation{Center for Astrophysics, Harvard \& Smithsonian, 60 Garden Street, Cambridge, MA 02138-1516, USA}

\begin{abstract}
We present images of IRAS~08235--4316 with the Dark Energy Camera Plane Survey (\textit{DECaPS}; spanning 0.398--1.034$\,\mu$m, at ${\sim}1''$ resolution) and the Submillimeter Array (\textit{SMA}; at 1.38\,mm/217\,GHz, at ${\sim}1.9''\times1.2''$ resolution), a YSO located in the Vela constellation near to the Puppis boundary, detected in a systematic search for new large/extended emission sources. 
The \textit{DECaPS} data show an asymmetric bi--polar morphology with a large angular extent of ${\sim}7.1''$ separated by a dark lane, characteristic of highly--inclined protoplanetary disks and less-evolved YSOs with outflows.
The \textit{SMA} data show an extended continuum structure along the optical dark lane with a smaller angular extent of ${\sim}4.6''$. 
The detected $^{12}$CO J=2--1 emission tentatively shows a velocity gradient along the position angle of the dark lane/millimeter continuum, that may trace rotating gas.
Additional $^{12}$CO emission is present which could trace infalling/outflowing gas, and/or a nearby gas cloud.
We estimate a distance to IRAS~08235--4316 of at least ${\sim}191\,$pc.
Supported by additional SED modelling, we infer IRAS~08235--4316 to be a newly discovered class~I YSO with an outflow, host to an embedded protoplanetary disk, with a large millimeter radius of ${\sim}440$\,au and dust mass ${\gtrsim}11\,M_\oplus$.
\end{abstract}


\section{Introduction} \label{sec:intro}
Planets and kilometer-sized planet-forming planetesimals are built-up from the small solids and molecular gas inherited from the interstellar medium, which combine together within gas- and dust-rich protoplanetary disks \citep{Williams11, Andrews18}. 
The statistical understanding of nearby protoplanetary disk demographics -- their typical luminosities and radial sizes -- has grown tremendously over the past decade, owing in large-part to millimeter surveys with the \textit{Submillimeter Array} (\textit{SMA}) and the \textit{Atacama Large Millimeter/sub-Millimeter Array} \citep[\textit{ALMA}; see e.g.,][]{Andrews13, Barenfeld16,Ansdell16,Pascucci16, Ansdell18, Cieza19, Williams19, Villenave21, Lovell21a, Manara23}.
Nevertheless, a \textit{complete} picture of even nearby star-forming region YSO population demographics remains lacking, whereby many protoplanetary disks are yet to be identified or characterised, and in particular highly-inclined sources which remain largely missing from the census \citep[i.e., due to their orientation, these can be up to 2--3 mag fainter in the K-band compared to stars of the same spectral type with less inclined disks;][]{Angelo2023}.

Whilst millimeter studies of disk properties are essential to characterise the material present in disks, detecting these around young stellar objects (YSOs) in the first instance has typically proven more efficient at optical and near-/mid-infrared wavelengths \citep[such as with \textit{2MASS}, \textit{WISE} and \textit{Spitzer} surveys and spectral energy distribution (SED) analysis, see for example][]{Evans03, Luhman12, Teixeira20}.
Another avenue for identifying and characterizing the largest, early-stage YSOs with disks, is the study of their reflection nebulae.
This promise has been evidenced recently by the small survey of well--known protoplanetary disks by \citet{Gupta23}, and further with the discovery of the giant edge-on disk IRAS~23077+6707 \citep[]{Berghea24, Monsch24}, as found in \textit{Pan-STARRS} optical images.
Indeed, utilizing available all-sky data in the near-, mid- and far-infrared is proving to be a viable tool for discovering new protoplanetary disks missed by targeted surveys \citep[see, e.g.,][]{Marquez2024}.

In this work we present resolved images of IRAS~08235--4316, identified via by-eye examination of \textit{DECaPS} optical scattered light images of a sample of sources photometrically selected from (extended source) \textit{2MASS} and \textit{AKARI} data.
We supplement our work with the first Submillimeter Array (\textit{SMA}) follow-up observations of IRAS~08235--4316 at 1.38\,mm (217\,GHz).
Our analysis suggests that IRAS~08235--4316 is a large ($\gtrsim$1000\,au extent), highly--inclined class~I YSO with an outflow, and an embedded protoplanetary disk.
We present in \S\ref{sec:discovery} the method adopted to find this source, in \S\ref{sec:data} all collected data and their reduction/calibration methods (where undertaken). 
In \S\ref{sec:analysis} we present our analysis of the source and discuss these observations in the context of the Vela star--forming region and the distance that we associate with IRAS~08235--4316.
We summarise our conclusions in section \S\ref{sec:conclusions}.

\section{Detection of IRAS~08235--4316} \label{sec:discovery}
Giant edge-on disks and YSOs with extended outflows can be readily resolved as bipolar nebulae in their optical scattered light emission, even with relatively low-resolution optical images, as demonstrated in the case of the largest known protoplanetary disks, for example IRAS~04158+2808 \citep{Glauser2008}, IRAS~18059--3211 \citep[`\textit{Gomez's Hamburger}'; see][]{Ruiz1987} and IRAS~23077+6707 \citep[`\textit{Dracula's Chivito}'; see][]{Berghea24, Monsch24}. 
As such, we considered the question: \textit{how rare do YSOs with comparable infrared properties present in available optical data?}
With the availability of optical images for the southern galactic plane \citep[i.e., in \textit{DECaPS;}][]{Saydjari2023} and the northern sky \citep[i.e., in the $3\pi$ \textit{Pan-STARRS} survey, cf.][]{Chambers2016} such a question can be considered systematically. 
However, owing to the immense data volumes of these surveys, investigating via by-eye-inspection of optical images necessarily requires us to restrict our search to a small sample of sources with properties consistent with known, large disks.

To build an initial sample, we cross-matched the \textit{AKARI} YSO catalogue of \citet{Toth2014} with the all-sky \textit{2MASS} extended source catalogue \citep{Jarrett2000, Skrutskie2006, 2MASSXSC}\footnote{We note that \citet{Toth2014} has been retracted from the literature owing to `critical errors in references and in tables', which as such, includes many falsely-identified galaxies and planetary nebulae (and may not be complete for nearby disks).
This catalog nevertheless is host to many thousands of far-infrared identified YSOs, and thus provides a list of sources from which we can search for optically-bright YSOs.}.
The cross-match was executed within {\tt topcat} \citep{topcat} using a search radius of $1''$.
The \textit{2MASS} extended source catalogue includes only sources with near-infrared emission spanning more than ${\gtrsim}4''$, thus only the very largest, near-infrared sources are present in the cross-match.
To limit the search, we placed a cut in K-band photometry $\rm{K}<11.5$ (restricting the search on only the brightest sources, but those with properties consistent with IRAS~23077+6707 and IRAS~18059--3211)\footnote{
These cuts necessarily mean that many protoplanetary disks will have been excluded. Whilst we successfully detected previously unidentified disks, the sample does not place robust statistical limits on the rarity with which these extended YSOs pervade the nearby Galaxy.}.
In total, the photometric-cut sample provided list of a few hundred target sources within the boundaries of the \textit{DECaPS} survey area.

\begin{figure}
    \centering
    \includegraphics[width=1.0\linewidth,clip, trim={0.25cm 0.2cm 0cm 0cm}]{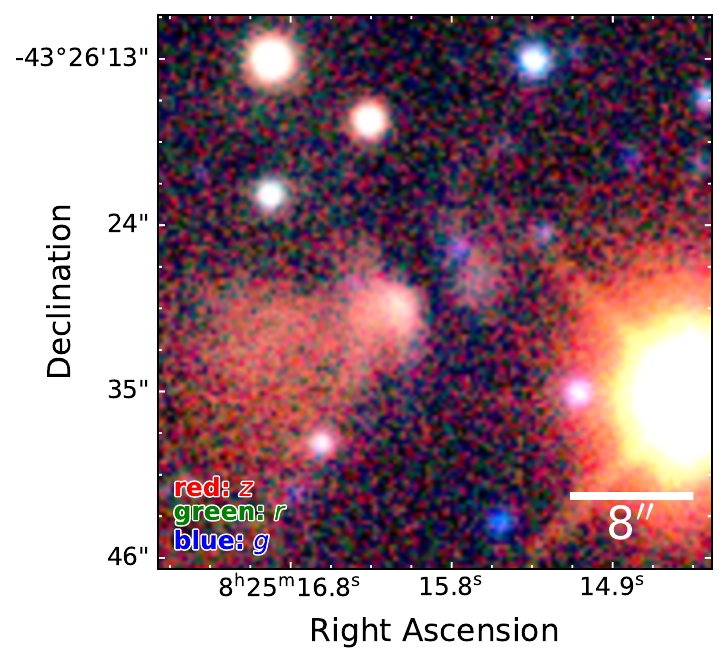}
    \caption{\textit{RGB image of IRAS~08235--4316 (`z', `r' and `g' filters respectively). A conical bright lobe, and an adjacent fainter, parallel lobe are separated by a dark central lane, as expected for highly-inclined embedded sources with outflows \citep[see e.g.,][]{Woitas2002, Delabrosse2024}. The $8''$ scale bar spans the approximate extent of the dark lane.}}
    \label{fig:LGC_RGB}
\end{figure}

Utilizing {\tt Aladin} \citep{Aladin}, we searched all target source locations within \textit{DECaPS} image data.
We identified a number of bipolar nebulae, including previously classified protoplanetary disk sources, and other sources which to our knowledge have not before been identified/studied (which we will present in future work).
The most promising candidate that we identified was IRAS~08235--4316 (see Fig.~\ref{fig:LGC_RGB}).
IRAS~08235--4316 is located in the Vela constellation, near to the boundary of the Puppis constellation \citep{Delporte1930}, a rich region of active star formation \citep[see for example][]{Murphy1991, Liseau1992, Yamaguchi1999, Massi2007}.
IRAS~08235--4316 is a few degrees south of the Vela `C' and `D' clouds \citep{ Massi1999, Massi2003} and likely owing to the complex, clustered environment with which it is co-spatial (additionally being located between the Vela Supernova Remnant at ${\sim}287\,$pc, see e.g., \citet{Large1968} and \citet{ Sushch2011}, and the more distant Puppis A Supernova remnant at ${\sim}1300\,$pc, see e.g.,  \citet{Milne1971} and \citet{Reynoso2017}), appears to have been over-looked by previous \textit{IRAS} surveys of YSOs \citep[e.g.,][]{Liseau1992}.

\section{Data collation and reduction} \label{sec:data}
\subsection{Dark Energy Camera Plane Survey (\textit{DECaPS})}
\textit{DECaPS} \citep[the Dark Energy Camera Plane Survey, see][]{Schlafly17, Saydjari22} is an optical--near-infrared southern galactic plane survey ($|b|{<}10^\circ$; $-124^\circ{<} l{<}6^\circ$) which uses the 4--meter Blanco Telescope in Chile with 5 filters ($grizY$; 398--1034\,nm) down to sub--20th magnitude, and ${\sim}1''$ seeing.
These properties make \textit{DECaPS} data ideal to hunt for faint, extended emission sources.
Having detected extended emission from IRAS~08235--4316 (see \S\ref{sec:intro} and \S\ref{sec:discovery}) we downloaded the available \textit{DECaPS} images for this source from the NOIRLab Astro Data Archive\footnote{\url{https://astroarchive.noirlab.edu/}} which provides pre-calibrated images from the \textit{DECaPS} archive via request.
We present in Fig.~\ref{fig:LGC_RGB} a \textit{DECaPS} red-green-blue (RGB) image using the $z$, $r$ and $g$ filters (i.e. wavelengths of 0.926\,$\mu$m, 0.642\,$\mu$m, and 0.473\,$\mu$m, respectively).

The \textit{DECaPS} data strongly resemble the scattered light emission of a highly-inclined class~I YSO with an outflow based on its bipolar morphology, and similarity to e.g., FS~Tau~B and DG~Tau~B \citep[see e.g.,][respectively]{Woitas2002, Delabrosse2024}.
Specifically at optical wavelengths, IRAS~08235--4316 is host to a bright fan-shaped/conical-shaped nebulosity (the outflow), adjacent to a dark lane where no scattered light is observed (where the outflow casts a shadow), and a faint, elongated conical region parallel to the dark lane (the lower emission surface).
We measure a scattered light extent on the brighter half of the nebula of ${\sim}7.1{\pm}1.0''$\footnote{We measure this from the maximum $3\sigma$ contour level extent of the source, with an associated error based on the data resolution, which we estimate from the seeing of ${\sim}1.0''$.}, and an approximate position angle along the dark lane of ${\approx}25^\circ$ (measured anti-clockwise from north), though due to the relatively low signal-to-noise (SNR) of the \textit{DECaPS} data, this is likely uncertain by at least several degrees. 

We measure peak--peak brightness ratios in the range ${\approx}2.5{-}5$ in the three \textit{DECaPS} filters.
Assuming that IRAS~08235--4316's scattered light emanates from a protoplanetary disk, this brightness ratio indicates it is likely inclined by 65--80$^\circ$ from the plane of the sky \citep[based on consistency with the models of][e.g., see their Fig.~4]{Watson07}, making it highly-inclined.
We note two caveats based on these \textit{DECaPS} data and our measurements. 
Firstly, there is a foreground star spatially co-located in the north of the fainter side of the nebula, which may be skewing the measurement of the brightness ratios and the optical scattered light extent of the fainter side (this appears as a blue-ish peak in Fig.~\ref{fig:LGC_RGB}).
Secondly, there is large-scale emission towards the south-east (emanating from the brighter side of the nebula, which has an outflow-like/fan morphology) which may also affect the measurement of the brightness ratios, the optical scattered light extent and the absolute value of the inclination. Comparison with models of earlier-stage YSOs \citep[e.g. those of][]{Gramajo2010} and the morphology of IRAS~08235--4316 indicate that the above inclination range should still provide a reasonable bound on the sky-projected inclination of the source, though higher-resolution optical scattered light data alongside detailed modelling are required improve on this estimate. 

\begin{figure*}
    \centering
    \includegraphics[width=1.0\textwidth, trim={0cm 0cm 0cm 0cm}]{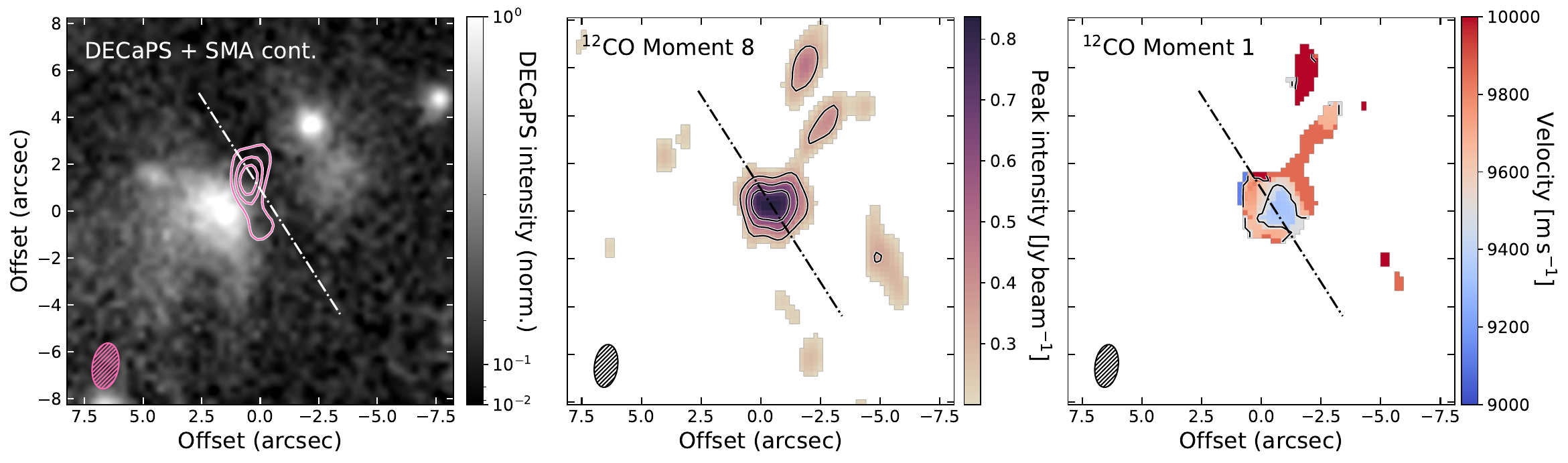}
    \caption{\textit{Left: \textit{SMA} millimeter contours over-plotted on the \textit{DECaPS} `z' filter image of IRAS~08235--4316. Center: $^{12}$CO moment~8 map. Right: $^{12}$CO moment~1 map. \textit{SMA} beams are shown in the bottom left of all panels. We over-plot a dash-dot line that represents the optical dark lane, which can be seen to pass through the peaks of the \textit{SMA} continuum and $^{12}$CO line data, and nearly-perpendicular to the plausible velocity gradient in the moment 1 map. In these maps, north is up, east is left.}}
    \label{fig:main}
\end{figure*}

\subsection{Submillimeter Array (\textit{SMA})}
\subsubsection{Data calibration}
The Submillimeter Array (\textit{SMA}) is an 8-dish (sub)millimeter interferometer based on Maunakea (\textit{Mauna a W\=akea}), Hawai'i \citep{Ho2004}.
IRAS~08235--4316 was observed on 16 Feb 2024 in project 2023B-S062 (PI: J. B. Lovell), for 139\,min on-source in the \textit{SMA}'s extended (`EXT') configuration with 7 antennas (1, 2, 3, 4, 5, 6 and 8) spanning projected baseline lengths of 16--179\,m to better understand the millimeter nature of this source.
The \textit{SMA} pointing center was centered on coordinates 08:25:15.94464, {-43}:26:28.35960 (RA, Dec., J2000).
Each receiver was tuned to a central local oscillator (LO) frequency of 217.0436\,GHz ($\lambda=1.38\,\mathrm{mm}$), with lower- and upper-sidebands spanning 202.025--214.025\,GHz and 222.025--234.025\,GHz respectively.
We observed J0747-331 and J0826-373 for gain calibration, 3c84 for bandpass calibration and Callisto for flux calibration ($413''$ separated from Jupiter during observations).
We converted the raw \textit{SMA} data to the \textit{Common Astronomy Software Applications} ({\tt CASA}) measurement set format (with a channel re--binning factor of 2) with {\tt pyuvdata} \citep{Hazelton+2017}.
We used the \textit{SMA} standard reduction script in {\tt CASA} version 6.4.1\footnote{The \textit{SMA} {\tt CASA} reduction script can be accessed via: \url{https://github.com/Smithsonian/sma-data-reduction}} to calibrate the measurement set.
During calibration we manually flagged all narrow interference spurs (that appeared in a small number of individual channels), and trimmed away 2.5\,\% of the overlapping ``guard-band'' regions between individual 2\,GHz SWARM correlator segments.

\subsubsection{Millimeter imaging and analysis}
\label{sect:obs_sma_imaging}
`Corrected' (calibrated table) data for IRAS~08235--4316 were extracted using the {\tt CASA mstransform} task with additional time-averaging of $20\,\mathrm{s}$. 
For analysing continuum emission, we re-bin all corrected data to 4 channels per $2\,$GHz spectral window, whereas for line emission analysis, we split these corrected data into the 12 separate \textit{SMA} spectral windows (with no channel averaging) and transform these to the local standard of rest frame (LSR).
We produce a continuum image and data cubes for the $^{12}$CO (J=2--1) and C$^{17}$O (J=2--1) lines with the {\tt CASA tclean} task, imaging to their respective $2\sigma$ thresholds (${\approx}0.6\,$mJy for the continuum, and ${\approx}150\,$mJy or ${\approx}110\,$mJy for the $^{12}$CO and C$^{17}$O lines respectively), adopting a `natural' weighting scheme for the continuum image (to maximise signal-to-noise), and a `briggs' weighting scheme (with a 0.5 robust parameter) for the CO-cubes (to provide improved spatial resolution of bright CO emission\footnote{We detected no significant emission from the C$^{17}$O line in either briggs robust 0.5 or natural-weighted cubes. From the natural-weighted cube, we estimate a 3$\sigma$ upper limit flux on this line of $370\,$mJy\,km\,s$^{-1}$ by measuring the dispersion in the flux of equal-width channel segments (over the measured line width of $^{12}$CO) in a C$^{17}$O line spectrum in quadrature with flux calibration uncertainty, extracted with an identical aperture to that used to measure $^{12}$CO (2--1) line flux, see later.}).
The resolution of the images are ${\sim}1.9''\times1.2''$ (continuum), ${\sim}1.8''\times1.0''$ ($^{12}$CO), and ${\sim}1.9''\times1.1''$ (C$^{17}$O). 
We present in Fig.~\ref{fig:main} the \textit{SMA} continuum contours over-plotted on the $z$ \textit{DECaPS} filter image, and moment maps of the $^{12}$CO J=2--1 image cube (described later).

We measure the total 1.38\,mm flux of IRAS~08235--4316 as $F_{\rm{1.38}}\,mm=9.8\pm2.0$\,mJy by fitting a Gaussian model to the central $8''\times5''$ region of the continuum image with the {\tt CASA imfit} task (where the reported error incorporates a standard \textit{SMA} 5\,\% flux calibration error in quadrature with the fitting uncertainty).
In addition, {\tt imfit} reports deconvolved major and minor axis extents of $4.6\pm1.1''$ and $1.2\pm0.4''$, and a position angle $11{\pm}7^\circ$ (measured anti-clockwise from north).
Whilst this position angle is less than that measured in the scattered light, these agree within their relative uncertainties.
The continuum image is slightly more radially extended towards the south versus the north, suggesting that the millimeter emission may be asymmetric. 
Higher resolution and/or deeper observations are required to verify this tentative finding.

To corroborate the continuum image-plane analysis, we also fit a 6-parameter `disk’ model to the measurement set visibilities with the {\tt CASA uvmodelfit} task.
We find best-fit parameters for the total flux of $8.6{\pm}1.3$\,mJy (where this error includes a 5\% flux calibration uncertainty in quadrature with the fitting uncertainty), RA and Decl. positional offsets of $0.24{\pm}0.10''$ and $-0.06{\pm}0.30''$ respectively, a major axis of $8.9{\pm}3.8''$, a major-minor axis ratio of $0.18{\pm}0.11$, and a position angle of $15{\pm}5^\circ$ (measured anti-clockwise from north).
The fitted parameters to the visibilities agree well with the fitted parameters to image data, and additionally, the disk model extent and position angle appear broadly consistent with the scattered light data. 
Nevertheless we highlight an issue in the recovery of all millimeter flux associated with IRAS~08235--4316 given the observational set up.
We use equation~A11 of \citet{Wilner1994} to approximate the 50th-percentile recoverable flux scale (RFS; $\Theta_{\frac{1}{2}}$) of the observations, i.e., 
\begin{equation} \label{eq:WilnerA11}
\Theta_{\frac{1}{2}}=8.6''\Big(\frac{\nu}{217\,\rm{GHz}}\Big)^{-1}  \Big(\frac{S_{\rm{min}}}{16\,\rm{m}}\Big)^{-1},
\end{equation}
for the observation frequency $\nu$ in GHz and the minimum projected baseline length $S_{\rm{min}}$ in m\footnote{This expression assumes a flat temperature structure disk. Instead, if the underlying temperature structure was relaxed to that of a Gaussian, the pre-factor in equation~\ref{eq:WilnerA11} alters from $8.6''$ to $7.0''$, which does not alter the conclusions we present in relation to flux recovery.}. 
We calculate values of $8.6''$ and $8.2''$ for the continuum and $^{12}$CO line observations respectively for the minimum projected baseline length of 16m, adopting either the \textit{SMA} LO frequency ($217.0436\,$GHz) or the $^{12}$CO J=2--1 transition frequency ($230.538\,$GHz). 
Due to the non-circular uv-coverage of the observations, these RFS estimates however fall to $5.4''$ and $5.2''$ respectively (along the perpendicular uv-baseline axis) at the continuum and $^{12}$CO frequencies.
As the scale size of the continuum emission RFS is comparable with the {\tt uvmodelfit} disk model extent, we have plausibly resolved out some fraction of the continuum emission. 
As such, this measurement of the total continuum flux is possibly underestimated.

\begin{figure}
    \centering
    \includegraphics[width=0.47\textwidth, trim={0cm 0cm 0cm 0cm}]{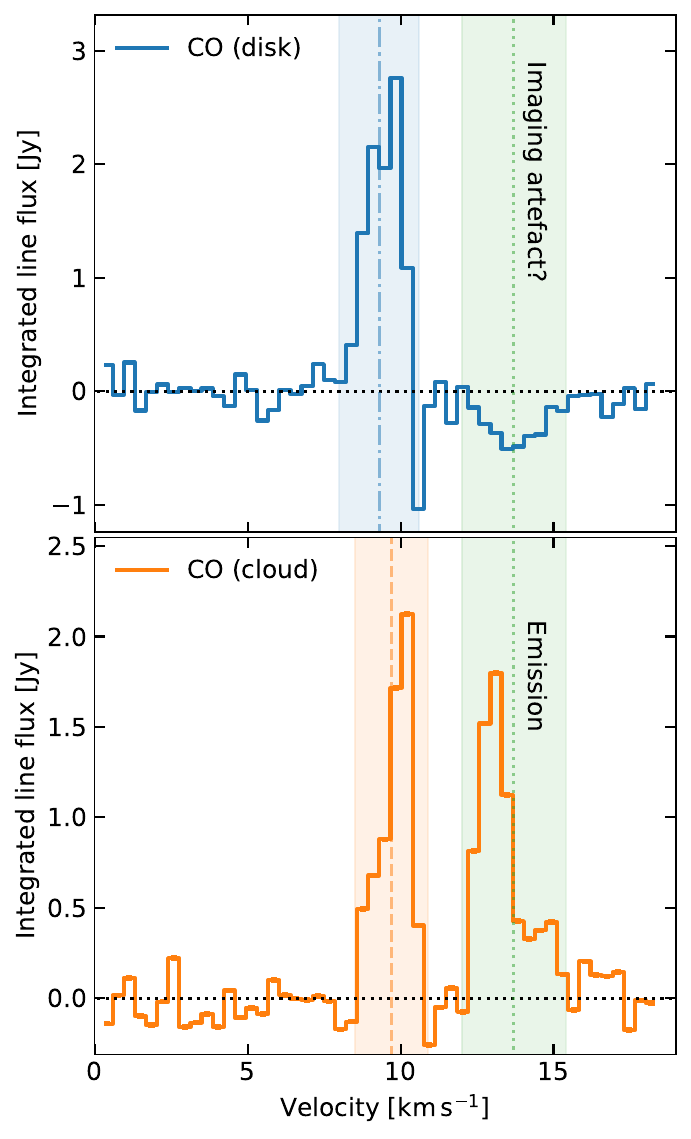}
    \caption{\textit{$^{12}$CO spectra for apertures centered on the central source (top) and the north-west cloud (bottom). We shade in the four regions associated as being distinct CO components.}}
    \label{fig:spectra}
\end{figure}

Inspecting the $^{12}$CO cube by-eye, we find evidence of two unique components.
To analyze the $^{12}$CO data, we produce two spectra with {\tt gofish} \citep{Teague2019_gofish}, one centered on the $^{12}$CO line peak at the central source location (with a circular aperture, and an outer radius of $2''$, which we call the `disk' spectrum) and one centered north-west of the central source (with a circular aperture, and an outer radius of $3.25''$, which we refer to as the `cloud' spectrum).
These two spectra are presented in  Fig.~\ref{fig:spectra} (while Fig.~\ref{fig:channels} shows channel maps over those fractions of the cube, in which the two spectra peak, respectively).
We first discuss the disk spectrum, from which we measure an integrated line flux of $3400{\pm}190\,$mJy\,km\,s$^{-1}$ (we estimate the uncertainty based on the quadrature sum of the \textit{SMA} flux calibration uncertainty in combination with the dispersion in the flux of equal-width line free channels in the spectrum). 
All velocities that we present are in the LSR frame.
The line center appears at ${\sim}9.3\,$km\,s$^{-1}$ and spans a velocity range 8.0--10.2\,km\,s$^{-1}$ for a line width of ${\sim}2.2\,$km\,s$^{-1}$.
We also find evidence of a slightly broader negative component, with a line center at ${\sim}13.7\,$km\,s$^{-1}$ (this feature is most plausibly an imaging artifact due to presence of extended emission, highlighted in green on the top plot of Fig.~\ref{fig:spectra}).

Utilizing {\tt bettermoments} \citep[][]{TeagueForeman-Mackey2018} we obtain moment 0 (integrated intensity), 1 (intensity weighted average velocity), and 8 (line peak intensity) maps, with clips of $2\sigma$, $2\sigma$ and $4.5\sigma$ on the moment 0, 8 and 1 maps respectively, smoothing parameters of $1\sigma$, $1\sigma$ and $0$ respectively, and include only channels in the frequency range $230.5300{-}230.5315$\,GHz, i.e., the frequency range over which $^{12}$CO \textit{emission} co-located with the disk spectrum is observed (i.e., the blue shaded region in the top panel of Fig.~\ref{fig:spectra}).
We present in Fig.~\ref{fig:main} the moment 1 and 8 maps, where we mask the moment 1 and 8 images for all pixels that exceed $0.2\,$Jy\,beam$^{-1}$ in the moment 8 map.
From the line peak map, it is evident that the brightest $^{12}$CO emission emanates from the same region where the continuum emission peaks.
Moreover, from the velocity map it is further evident that blue-- and red--shifted emission is present co-spatially with the continuum emission (and dark lane), which hints at a plausible velocity gradient parallel to the dark lane/millimeter position angle possibly indicating the presence of gas in Keplerian rotation (with a velocity width and line-center consistent with CO-line observations of protoplanetary disks).
Owing to the low resolution of the data, PV diagrams provide no further insight into this emission.
We find no evidence of $^{12}$CO emission in the southeast where the optical scattered light presents the fan-shaped emission. We emphasize that this may not be present simply due to the sensitivity limitations, though discuss this morphology further below.

Intriguing in the $^{12}$CO cube are additional features that are spatially and kinematically distinct from the central source $^{12}$CO (which we refer to as the `cloud').
Firstly, there is a bright, asymmetric lobe to the north-west of IRAS~08235--4316 which spans a larger area than the $^{12}$CO observed towards the central source (physically offset by $3''$ northwards, $5''$ westwards) with a line center of ${\sim}9.7\,$km\,s$^{-1}$ (receding slightly \textit{faster} than the line center of the central component).
As seen in both Figs.~\ref{fig:main} and \ref{fig:channels}, this component is spatially extended from the center, and with a velocity gradient from 8.5--10.9\,km\,s$^{-1}$ that approaches that of the disk spectrum (consistent with gas either being ejected from, or falling on to a disk).
Since this component spans a spatial scale comparable with the RFS we do not report its total flux, but note this has a peak brightness approximately half that of the central source.
We speculate that this component could be tracing an infalling accretion streamer/a late-stage infall event, or a molecular wind/outflow \citep[such as those discussed in, for example][]{Alves2019, Pineda2020, Kuffmeier2021, Garufi2022, Gupta23, Pascucci2023, Pineda2023, Hanawa2024, Hales2024}.
Comparison with other YSOs observed at higher-resolution may illuminate our understanding of IRAS~08235-4316. For example, in the case of DG~Tau~B (which is morphologically very similar in scattered light emission) ALMA data shows CO to be significantly brighter and more extended on the opposite side of the source as its scattered light fan-shaped emission \citep[for which gas is known to be part of an outflow;][]{Delabrosse2024}. The structural similarity between IRAS~08235-4316 and DG~Tau~B in their scattered light emission and CO morphology and kinematics may therefore lend support to this extended CO component tracing an outflow.
Secondly, there is emission red-shifted to ${\sim}13.5$\,km\,s$^{-1}$, at an identical velocity and line-width as seen in the imaging artifact.
Most plausibly, this component is tracing resolved-out emission from a nearby CO-rich gas cloud (a fraction of which could be infalling on to the disk, evidenced by the blue-shifted velocity component). 
Combined, these distinct features demonstrate that there is a significant reservoir of CO in the vicinity of IRAS~08235--4316, distinct from that present within the region we associate with a disk. 
We leave as an open question how this additional gas reservoir may be interacting with the disk.

\begin{table}[]
    \centering
    \caption{\textit{Collated flux density values for IRAS~08235--4316, from near-infrared to millimeter wavelengths. References are as follows: A--\citet{Skrutskie2006, 2MASSXSC}; B--\citet{Skrutskie2006, allwise_data, allwise_2doi2}; C--\citet{Abrahamyan2015}; D--\citet{SEIP_spitzer}; E--\citet{AKARI2010} \& \citet{Toth2014}. Errors denoted with $^*$ were assumed at the level of 20\% given these have not been published elsewhere. Instrument data denoted with $^{+}$ were not modelled in the SED fit.}}
    \begin{tabular}{l|c|c|c|c}
    \hline
    \hline
      Instrument & wavelength & Flux & error & Ref. \\
       & [$\mu$m] & [mJy] & [mJy] &  \\
     \hline
    \textit{2MASS} J & 1.24 & 5.40 & 0.12 & A \\
    \textit{2MASS} H & 1.66 & 16.73 & 0.20 & A \\
    \textit{2MASS} K & 2.16 & 27.92 & 0.247 & A \\
    \textit{WISE} W1$^{+}$ & 3.3 & 10.94 & 0.3 & B \\
    \textit{WISE} W2$^{+}$ & 4.7 & 23.58 & 0.05 & B \\
    \textit{WISE} W3$^{+}$ & 12.0 & 34.92 & 0.09 & B \\
    \textit{IRAS} F12 & 12.0 & 260 & 50$^*$ & C \\ 
    \textit{WISE} W4 & 22.0 & 287.0 & 2.1 & B \\
    \textit{Spitzer} MIPS & 23.7 & 275.2 & 0.275 & D \\
    \textit{IRAS} F25 & 25 & 240 & 50$^*$ & C \\ 
    \textit{IRAS} F60 & 60 & 2400 & 500$^*$ & C \\
    \textit{AKARI} S90$^{+}$ & 90 & 3670 & 100 & E \\
    \textit{IRAS} F100 & 100 & 8200 & 1600$^*$ & C \\ 
    \textit{AKARI} S140 & 140 & 7700 & 700 & E \\
    \textit{AKARI} S160 & 160 & 5600 & 900 & E \\
    \textit{SMA} & 1380 & 8.6 & 1.3 & This work\\ 
    \hline
    \end{tabular}
    \label{tab:fluxes}
\end{table}

\subsection{Spectral Energy Distribution}
The resolved \textit{DECaPS} and \textit{SMA} images together strongly imply the presence of a highly inclined YSO.
To verify this at other wavelengths, we collate available photometry for IRAS~08235--4316.
We present in Table~\ref{tab:fluxes} the tabulated fluxes extracted for IRAS~08235--4316.
From the \textit{2MASS} K-band and \textit{Spitzer} MIPS photometry we measure IRAS~08235--4316's mid-infrared spectral index as 0.6, suggesting this to be a class~0/I YSO, based on 
\begin{equation}
    \alpha_{\rm{IR}} = \frac{\partial \log{\lambda F_\lambda}}{\partial \log{\lambda}},
\end{equation}
\citep[see e.g.,][]{Lada87, Adams87, Williams11}.
It has been noted previously in the literature however that highly-inclined YSOs likely have their spectral indices over-estimated due to preferential extinction of their near-infrared emission \citep[see also][]{Greene1994}.
Consequently, we infer this source as more likely a later stage class~I YSO. 

\begin{figure}
    \centering
    \includegraphics[width=1.0\linewidth, clip, trim={0.25cm 0cm 1.2cm 1.2cm}]{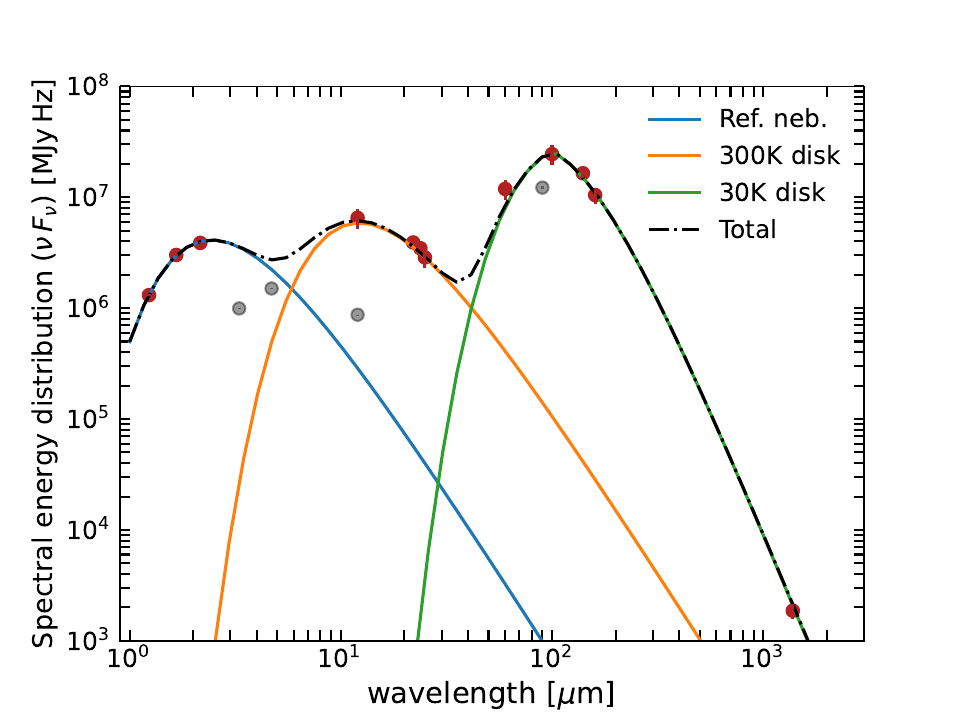}
    \caption{\textit{$\nu\,F_\nu$ versus wavelength, presenting all values from Table~\ref{tab:fluxes}. We plot the individual and combined model spectra of three blackbodies profiles (with temperatures of 1500\,K; blue, 300\,K; amber, and 30\,K; green). Four (of 16) data points are not well-described by the composite (marked in grey), whereas the other twelve data points are well described by the simple triple-blackbody distribution.}}
    \label{fig:SED}
\end{figure}

On Fig.\ref{fig:SED} we present the complete spectral energy distribution (SED) for the collated data, and over-plot three blackbody spectra with temperatures of 1500\,K (blue, a hot stellar-like component), 300\,K (amber, a warm inner component) and 30\,K (green, a cold outer disk-plus-envelope-like component), with respective emitting areas expected for a young star, host to such inner warm and outer cold components.
Whilst 4/16 data points lie off the total composite spectrum (which could indicate that the reported fluxes in some surveys do not fully account for IRAS~08235--4316 being resolved), the rest of the data are well described by this composite spectrum.
The four offset data points are \textit{WISE} W1, W2 and W3, and \textit{AKARI} S90. For both \textit{WISE} and \textit{AKARI}, these filters are the highest-resolution, which may explain why these are systematically lower than the simple model would predict. 
In the case of W1 and W2 however we note that the difference could also be due to the simple model neglecting extinction.

For the colder component, we include a blackbody modification term that steepens the profile long-wards of $100\,\mu$m by $\propto \nu^{2+\beta}$ for $\beta=1.75$ \citep[readily done for optically thin disks, see e.g.,][]{WilliamsAndrews2006, Kennedy14}\footnote{The blackbody spectra were modelled purely by-eye and thus we place no estimate on the uncertainties of the emission surface areas or temperatures; the exercise we adopted here was conducted simply to investigate whether the flux densities were consistent with that anticipated for a highly inclined class~I YSO.}. 
Typically, protoplanetary disks have values of $\beta\leq1.0$, whereas the by-eye fit prefers a value closer to the ISM \citep{Testi2014}; which further suggests the millimeter emission may be dominated by emission from an envelope rather than an embedded protoplanetary disk, or instead that at these long wavelengths, the outer disk is optically thin, with a value in the upper-range of those measured for young protoplanetary disks \citep[see e.g.,][]{Andrews2007, Ribas2017}\footnote{The SED behaviour from the far-infrared to millimeter is agnostic to the possibility that continuum flux has been resolved out. We verified this conclusion by modelling the SED with twice the \textit{SMA} 1.38\,mm flux value which still requires a large $\beta=1.5$.}.
Given the scattered light emission morphology, it seems more likely that this emission is dominated by smaller grains in an envelope.
Better estimation of the optical depth of the large grains traced by the millimeter emission will require new centimeter-radio and sub-millimeter data (and shorter baseline millimeter interferometry data to rule-out the possibility that emission has been resolved-out by the \textit{SMA} observations we present here).

A detailed comparison with the edge-on/highly-inclined disk SED models presented by \citet{Angelo2023} is challenged by the possibility that IRAS~08235--4316 is an earlier-stage YSO and thus host to additional emission (e.g., from an extended envelope).
For example, in all the \citet{Angelo2023} models, their SED $1\,\mu$m emission is brighter than both the $10\,\mu$m and $100\,\mu$m emission, whereas the data we present for IRAS~08235--4316 shows the opposite (which could result from an extended envelope preferentially dimming the stellar component versus the mid- and far-infrared emission from larger grains).
Overall our analysis of the SED corroborates both the \textit{DECaPS} image and millimeter data, in that this spectrum analysis further suggests that IRAS~08235--4316 is indeed a class~I YSO, and likely host to a remnant protostellar envelope/outflow.

\section{Analysis and Discussion} \label{sec:analysis}
\subsection{A new large, highly-inclined class~I YSO with an outflow in Vela}
Providing first estimates on source parameters, such as its mass and size both require knowledge of the source distance.
As noted by \citet{Perrin2006} in the case of the highly-inclined YSO PDS~144 `\textit{The distance to PDS 144 remains frustratingly uncertain}'.
We conclude identically; given its edge-on/highly-inclined nature which obscures the central star, and thus lack of \textit{Hipparcos}-/\textit{Gaia}-based stellar-parallax, the distance to IRAS~08235--4316 is likewise frustratingly uncertain.

In this work we resort to estimating IRAS~08235--4316's distance via an analysis of the line-of-sight dust extinction \citep[utilizing the \textit{Gaia}-based 3D dust extinction maps of][which provide parsec-scale resolution out to $1.25\,$kpc distances]{Edenhofer24} on the assumption that IRAS~08235--4316 is associated with a dust-rich region along its line-of-sight.
Placing an annulus around IRAS~08235--4316 with a $1^\circ$ radius, we measure a range of dust extinction density peaks along the line-of-sight (LOS) at distances of $191\pm7$pc, $462\pm11$pc, $747\pm14$pc and $881\pm19$pc.

\begin{figure}
    \centering
    \includegraphics[width=1.0\linewidth]{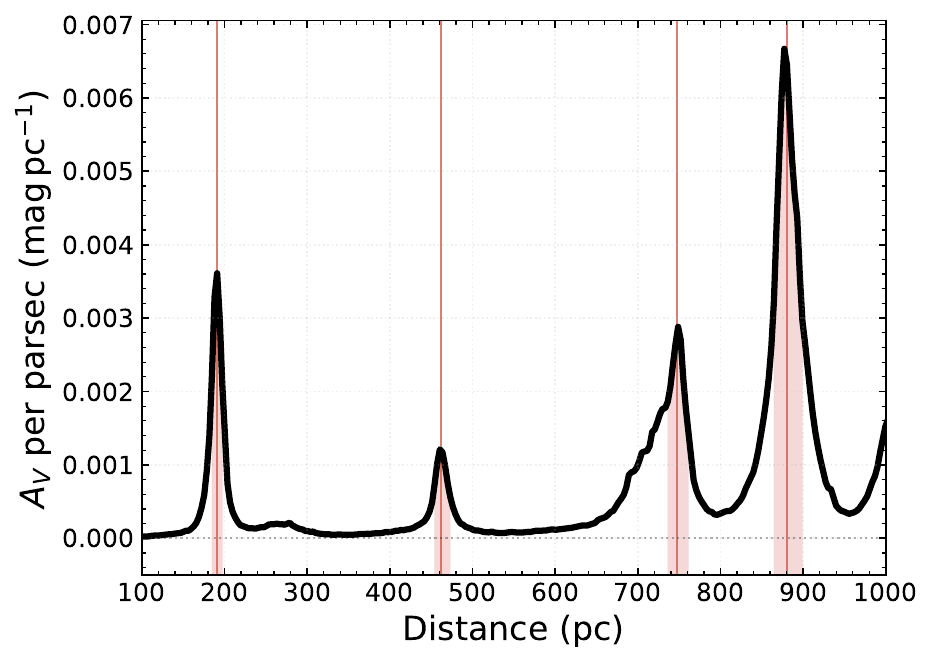}
    \caption{\textit{Dust extinction as a function of distance along the line-of-sight towards IRAS~08235-4316, based on the 3D dust extinction maps from \citet{Edenhofer24}. The data were extracted within a cone of radius $1^\circ$ around IRAS~08235-4316's sky position. There are four strong peaks associated with this direction, located at $191\pm7$pc, $462\pm11$pc, $747\pm14$pc and $881\pm19$pc (obtained by simple Gaussian fits).}}
    \label{fig:dist}
\end{figure}

We present in Fig.~\ref{fig:dist} the resulting posterior mean map of the dust extinction as a function of distance (with red lines and shaded widths showing the mean and standard deviations of the four peaks respectively, obtained by fitting 1D Gaussian profiles to the individual peaks).
On the assumption that IRAS~08235--4316 is associated with the closest of these dusty regions (as this provides the most conservative estimate for all distance-derived YSO-parameters), we continue the analysis with a distance estimate of $191\pm7$pc for IRAS~08235-4316\footnote{Since the subsequent analyses relies on this adopted distance, we note that radius/extent estimates scale linearly with distance, and mass estimates scale with the square of the distance.}.
Inspection of the wider field nearby to IRAS~08235-4316 shows this source to be near to the edge of a much larger dark cloud (to the west), which may lend support to the association between IRAS~08235-4316 and this closest dark cloud along our line-of-sight.

With an assumed distance of 191\,pc, we can estimate the radius and mass.
From the scattered light and millimeter emission we measured extents of $7.1{\pm}1.0''$ and $4.6{\pm}1.1''$ respectively. From these diameters, we quantify the plausible scattered light and millimeter radii as $680\,{\pm}190\,$au and $440{\pm}110\,$au respectively (i.e., extents of 1360\,au and 880\,au).
If dominated by disk emission, these values would place IRAS~08235-4316's (millimeter) radius in the upper-end of the typical protoplanetary disk radii distribution \citep[see e.g. the tabulated values in][]{Manara23}, and larger than resolved rings observed in the Disk Substructures at High Angular Resolution Project \citep[DSHARP; see both][]{Andrews18, Huang18}. 
In comparison to other well-studied highly-inclined/edge-on disks, IRAS~08235-4316 presents slightly above the mean of the population, i.e. in comparison to higher-resolution \textit{ALMA} observations presented in \citet{Villenave2020}, and comparable resolution \textit{SMA} observations of e.g., the `Butterfly Star' (IRAS 04302+2247), the `Flying Saucer' (2MASS J16281370-2431391), `Gomez's Hamburger' (IRAS~18059-3211), and IRAS~23077+6707 \citep[see][]{Wolf2008, Guilloteau2016,Teague2020,Monsch24}.
Higher resolution, higher signal-to-noise observations are however needed to distinguish the extent to which this emission traces a disk or envelope component, and place stricter bounds on the estimated radius, and determine the radial distribution of matter.

Even with accurate stellar distances, disk masses are notoriously uncertain owing to optical depth, temperature and grain-size distribution uncertainties \citep[see for example][]{Liu2022, Miotello2023,Kaeufer2023, Stapper2024}.
Nevertheless, we can place an estimate on both the dust mass and total mass ($100\times$ the dust mass) from the continuum flux measurement, utilizing 
\begin{equation}
    M_{\rm{dust}} [M_{\oplus}] = \frac{F_{\nu}d^2}{\kappa_{\nu}B_{\nu}(T_{\rm{dust}})},
\end{equation}
for the distance $d$ (in pc), flux $F_\nu$ (in mJy) with an assumed dust temperature of 20\,K \citep[the median used to measure disk masses in Taurus, see][]{Andrews05} and a dust-grain opacity $\kappa_{\nu}$ of $10\,\rm{cm}^{2}\,\rm{g}^{-1}$ (at 1000\,GHz) with an opacity power-law index $\beta=1$ \citep[e.g.,][]{Hildebrand83, Beckwith90}. 
We measure the dust mass via this method as $\approx11\,M_\oplus$, and assuming a dust-to-gas ratio of 1:100, a total mass of $\approx1100\,M_\oplus$ (or $\sim 3.5$ Jupiter masses).
This estimate is highly uncertain for the reasons outlined above, but is typical for many YSOs as surveyed thus far in the millimeter, with sufficient mass to enable giant planet formation.
To improve on this mass estimate, longer wavelength continuum observations, or optically thin gas tracer observations are required \citep[see e.g.,][]{Miotello2016, Anderson2019, Trapman2022, Stapper2024}.

We cannot rule out the possibility that IRAS~08235--4316 is associated with one of the more distant dust extinction peaks \citep[which are likely connected to emission associated with Trumpler~10 and/or Vela~OB2, see Fig.11 of][]{deZeuw99}.
The greater distance of ${\sim}462\,$pc would be more consistent with that measured to Vela~OB2 \citep[based on the \textit{Hipparcos} stellar distance ladder presented by][with a mean associated distance of 410\,pc]{deZeuw99}, and various clusters and associations within the Vela complex \citep[based on the \textit{Gaia} stellar distance ladder presented by][with a mean distance of 354\,pc]{Kerr2023}.
We are unable to compare this assumed value of 191\,pc with \textit{Gaia}-based kinematic membership assessments of ${\sim}$191\,pc stars in the vicinity of Vela/IRAS~08235--4316 however due to sample cuts in existing stellar catalogs \citep[typically only stars more distant than 250\,pc have been included in these, e.g.,][which have systematically sampled out more nearby stars that IRAS~08235--4316 could be associated with]{Armstrong2018,Cantat2019}.
We leave to future work further analysis on this distance estimate (and moreover, its plausible formation history, given IRAS~08235--4316's proximity to the Vela Supernova Remnant).

\section{Summary and Conclusions} \label{sec:conclusions}
We have presented the first optical scattered light \textit{DECaPS} and millimeter \textit{SMA} images of IRAS~08235--4316, an extended source located near the boundary of the Vela and Puppis constellations. 
IRAS~08235--4316 was found by-eye as part of a systematic hunt for edge-on protoplanetary disks via their scattered light nebula.
We found: 

\begin{enumerate}[leftmargin=5.0mm]
\item In the \textit{DECaPS} data, IRAS~08235--4316 presents an asymmetric bi--polar morphology separated by a dark--lane, characteristic of highly--inclined protoplanetary disks and earlier-stage YSOs with outflows, with a large angular extent of ${\sim}7.1''$.
In the \textit{SMA} data, IRAS~08235--4316 presents a continuum structure which is coincident with the optical dark lane, with an angular extent of ${\sim}4.6''$, and is co-located with $^{12}$CO J=2--1 gas emission which has a plausible velocity gradient parallel with the dark lane (suggesting the gas is in rotation).
We collated available photometry for IRAS~08235--4316 and presented its spectral energy distribution, which shows consistency with the expected profile of a YSO embedded in an envelope, or an edge-on circumstellar disk.
Overall, we infer that IRAS~08235--4316 is a newly discovered, highly-inclined, class~I YSO with an outflow. 

\item We estimated four distances to IRAS~08235--4316.
Assuming the lowest of these estimates, ${\sim}191\,$pc, we infer that IRAS~08235-4316 has a millimeter radius of $440$\,au, a scattered light extent of 1360\,au, and a dust mass ${\gtrsim}11\,M_\oplus$ (with plausibly much higher values if instead the source resides at one of the larger distance estimates).

\item Additional $^{12}$CO emission is present, which traces a distinct gas component. Whilst uncertain, this may trace either outflowing gas, or gas infalling from an external cloud as a streamer, present only on the fainter side of IRAS~08235--4316 in scattered light.
\end{enumerate}

On the basis that IRAS~08235--4316 is a large edge-on YSO, following the sandwich-inspired unofficial names designated to IRAS~18059-3211 (\textit{Gomez's Hamburger}) and IRAS~23077+6707 (\textit{Dracula's Chivito}); GKK proposes an unofficial name for IRAS~08235--4316: ``\textit{Lovell's Grilled Cheese}''.

\section{Software and third party data repository citations} \label{sec:cite}
\software{{\texttt{APLpy} \citep{RobitailleBressert2012}, \tt astropy} \citep{2013A&A...558A..33A,2018AJ....156..123A}, {\tt aladin} \citep{Aladin}, {\tt bettermoments} \citep{TeagueForeman-Mackey2018}, {\tt CASA} \citep{McMullin+2007}, {\tt gofish} \citep{Teague2019_gofish}, {\tt pyuvdata} \citep{Hazelton+2017}, {\tt topcat} \citep{topcat}.}
All data and code used in this project have been uploaded to JBL's \textit{github} at \url{https://github.com/astroJLovell/LovellsGrilledCheese} and are stored on Zenodo at \url{https://doi.org/10.5281/zenodo.14163373} \citep{astroJLovell_2024_LGC}.

\appendix
\section{$^{12}$CO channel maps}
\begin{figure*}
    \centering
    \includegraphics[width=0.85\linewidth]{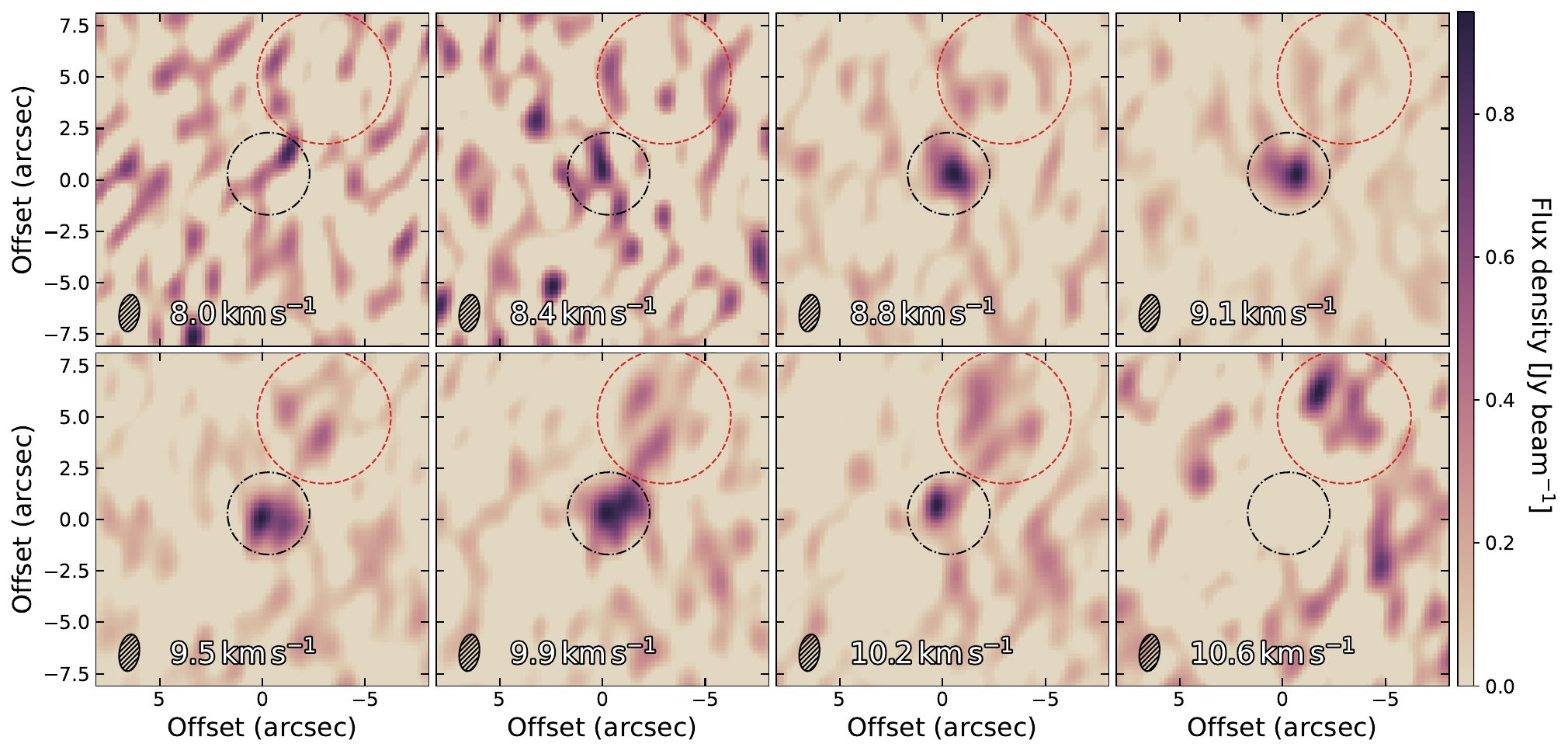}
    \caption{\textit{$^{12}$CO channel maps, from 8.0--10.6\,km\,s$^{-1}$. The central component is highlighted by a black dash-dot ring. The spatially separated (at $-3''$, $5''$) northern extended component is highlighted by a red dash ring. These rings represent the same apertures from which we measure both $^{12}$CO flux spectra. The SMA beam is shown in the lower-left of all panels. In these maps, north is up, east is left.}}
    \label{fig:channels}
\end{figure*}

We present in Fig.~\ref{fig:channels} the $^{12}$CO channel maps from 8.4--10.9\,km\,s$^{-1}$ to demonstrate the multiple components of CO gas traced.
We over-plot circular apertures which represent the two regions over which we extract flux spectra in Fig.~\ref{fig:spectra}, one co-located with the central source, the other to the northwest where the additional (cloud) emission peaks.
Whilst the SNR of the respective peaks in these separate channel maps is relatively low, the source in the central aperture appears partially spatially resolved, peaking from 8.4--9.1\,km\,s$^{-1}$ on its west side (right), and peaking from 9.5--10.2\,km\,s$^{-1}$ on its east side (left). This emission gives rise to the tentative velocity gradient noted in the discussion of the moment-1 map, which could be evidence that the $^{12}$CO emission is tracing the outer regions of a rotating (embedded) protoplanetary disk.

\begin{acknowledgments}
\break
\textit{\large{Acknowledgments:}}
JBL acknowledges the Smithsonian Institute for funding via a Submillimeter Array (\textit{SMA}) Fellowship, and Mackenzie Whitaker for artistic input to the choice of colormaps used in Figs.~\ref{fig:main} and \ref{fig:channels}.
KM was supported by NASA grants GO8-19015X, TM9-20001X, GO7-18017X, HST-GO-15326 and JWST-GO-1905. 
GE acknowledges that support for this work was provided by the German Academic Scholarship Foundation in the form of a PhD scholarship (``Promotionsstipendium der Studienstiftung des Deutschen Volkes'').

The Submillimeter Array is a joint project between the Smithsonian Astrophysical Observatory and the Academia Sinica Institute of Astronomy and Astrophysics and is funded by the Smithsonian Institution and the Academia Sinica. 
The authors wish to recognize and acknowledge the very significant cultural role and reverence that the summit of Maunakea has always had within the indigenous Hawaiian community, where the Submillimeter Array (\textit{SMA}) is located. 
We are most fortunate to have the opportunity to conduct observations from this mountain. We further acknowledge the operational staff and scientists involved in the collection of data presented here. 
The \textit{SMA} data used here is from project 2023B-S062 and can be accessed via the Radio Telescope Data Center (RTDC) at \url{https://lweb.cfa.harvard.edu/cgi-bin/sma/smaarch.pl} after these have elapsed their proprietary access periods. 

This project used data obtained with the Dark Energy Camera (DECam), which was constructed by the Dark Energy Survey (DES) collaboration. Funding for the DES Projects has been provided by the DOE and NSF (USA), MISE (Spain), STFC (UK), HEFCE (UK), NCSA (UIUC), KICP (U. Chicago), CCAPP (Ohio State), MIFPA (Texas A\&M), CNPQ, FAPERJ, FINEP (Brazil), MINECO (Spain), DFG (Germany) and the collaborating institutions in the Dark Energy Survey, which are Argonne Lab, UC Santa Cruz, University of Cambridge, CIEMAT-Madrid, University of Chicago, University College London, DES-Brazil Consortium, University of Edinburgh, ETH Z{\"u}rich, Fermilab, University of Illinois, ICE (IEEC-CSIC), IFAE Barcelona, Lawrence Berkeley Lab, LMU M{\"u}nchen and the associated Excellence Cluster Universe, University of Michigan, NOAO, University of Nottingham, Ohio State University, OzDES Membership Consortium, University of Pennsylvania, University of Portsmouth, SLAC National Lab, Stanford University, University of Sussex, and Texas A\&M University.

This research uses services or data provided by the Astro Data Archive at NSF's NOIRLab. NOIRLab is operated by the Association of Universities for Research in Astronomy (AURA), Inc. under a cooperative agreement with the National Science Foundation.
This research has made use of ``Aladin sky atlas'' developed at CDS, Strasbourg Observatory, France.

This publication makes use of data products from the Two Micron All Sky Survey, which is a joint project of the University of Massachusetts and the Infrared Processing and Analysis Center/California Institute of Technology, funded by the National Aeronautics and Space Administration and the National Science Foundation \citep[10.26131/IRSA97][]{2MASSXSC}.

This publication makes use of data products from the Wide-field Infrared Survey Explorer, which is a joint project of the University of California, Los Angeles, and the Jet Propulsion Laboratory/California Institute of Technology, funded by the National Aeronautics and Space Administration \citep[10.26131/IRSA142;][]{allwise_2doi2}.

The Infrared Astronomical Satellite (IRAS) was a joint project of the US, UK and the Netherlands.
This research is based on observations with \textit{AKARI}, a JAXA project with the participation of ESA.

\end{acknowledgments}

\vspace{5mm}
\facilities{Smithsonian Astrophysical Observatory (SAO)/Academia Sinica SubMillimeter Array (\textit{SMA}) at Mauna Kea Observatory, NSF NOIRLab 4m Victor M. Blanco Telescope at NSF Cerro Tololo Inter-American Observatory (CTIO), The 2 (Two) Micron All Sky Survey (\textit{2MASS}) 1.3m Telescope at Cerro Tololo Inter-American Observatory (CTIO), NASA 0.4m Wide-field Infrared Survey Explorer (\textit{WISE}) Satellite Mission, Japan Aerospace eXploration Agency (JAXA) \textit{AKARI} Satellite Mission (formerly ASTRO-F), NASA/Netherlands Institute for Space Research (NIVR)/Science and Engineering Research Council (SERC) 0.57m InfraRed Astronomical Satellite (IRAS) Satellite Mission}

\bibliography{sample631}{}
\bibliographystyle{aasjournal}

\end{document}